\documentclass[prl,twocolumn,showpacs,superscriptaddress,nofootinbib,floatfix,10pt]{revtex4-1} 
\usepackage[utf8]{inputenc}
\usepackage{color}
\usepackage{amsfonts,amsmath,amssymb}
\usepackage{longtable,booktabs}
\usepackage{graphicx}
\usepackage{bm}
\usepackage{hyperref}
\usepackage{color}
\usepackage{epstopdf}
\usepackage{mathtools}

\newcommand{\re}{\text{Re}\,}
\newcommand{\im}{\text{Im}\,}

\begin{document}

\title{Explaining the Many Threshold Structures in the Heavy-Quark Hadron Spectrum}

\author{Xiang-Kun Dong}
\author{Feng-Kun Guo}\email[E-mail address: ]{fkguo@itp.ac.cn}
\affiliation{CAS Key Laboratory of Theoretical Physics, Institute of Theoretical Physics, Chinese Academy of Sciences, Beijing 100190, China}
\affiliation{School of Physical Sciences, University of Chinese Academy of Sciences, Beijing 100049, China
}

\author{Bing-Song Zou}
\affiliation{CAS Key Laboratory of Theoretical Physics, Institute of Theoretical Physics, Chinese Academy of Sciences, Beijing 100190, China}
\affiliation{School of Physical Sciences, University of Chinese Academy of Sciences, Beijing 100049, China
}
\affiliation{School of Physics, Central South University, Changsha 410083, China}

\begin{abstract}

Tremendous progress has been made experimentally in the hadron spectrum containing heavy quarks in the last two decades. It is surprising that many resonant structures are around thresholds of a pair of heavy hadrons.
There should be a threshold cusp at any $S$-wave threshold.
By constructing a nonrelativistic effective field theory with open channels, we discuss the generalities of threshold behavior, and  offer an explanation of the abundance of near-threshold peaks in the heavy quarkonium regime.
We show that the threshold cusp can show up as a peak only for channels with attractive interaction, and the width of the cusp is inversely proportional to the reduced mass relevant for the threshold. 
We argue that there should be threshold structures at any threshold of a pair of heavy-quark and heavy-antiquark hadrons, which have attractive interaction at threshold, in the invariant mass distribution of a heavy quarkonium and light hadrons that couple to that open-flavor hadron pair. The structure becomes more pronounced if there is a near-threshold pole.
Predictions of the possible pairs are also given for the ground state heavy hadrons.
Precisely measuring the threshold structures will play an important role in revealing the heavy-hadron interactions, and thus understanding the puzzling hidden-charm and hidden-bottom structures.
\end{abstract}

\date{\today}

\maketitle

\vspace{2cm}

{\it Introduction.}---Quantum chromodynamics, the fundamental theory of the strong interaction, is nonperturbative at low energies. As a result, the low-energy strong interaction is notoriously difficult to be tackled with, and so far the mechanism for the color confinement, i.e., all quarks and gluons are confined inside color neutral hadrons, remains the most challenging problem in the standard model. 
As a manifestation of color confinement, how the spectrum of hadrons emerges from the underlying strong dynamics remains mysterious.

One of the most puzzling issues in hadron spectroscopy is how the plethora of resonant structures observed since 2003 can be understood.
Most of them were observed in the invariant mass of hadrons containing heavy quarks. The meson(-like) structures in the heavy quarkonium mass region are called the $XYZ$ states to emphasize that the internal structure and why they appear in the mass spectrum have not been understood (see Refs.~\cite{Chen:2016qju,Esposito:2016noz,Hosaka:2016pey,Richard:2016eis,Lebed:2016hpi,Guo:2017jvc,Ali:2017jda,Olsen:2017bmm,Kou:2018nap,Cerri:2018ypt,Liu:2019zoy,Brambilla:2019esw,Guo:2019twa,Yang:2020atz} for recent reviews).

One salient feature of the new resonant hadron structures is that many of them have masses around the thresholds of a pair of hadrons. To name a few famous examples, let us mention the $X(3872)$~\cite{Choi:2003ue} and $Z_c(3900)^\pm$~\cite{Ablikim:2013mio,Liu:2013dau,Ablikim:2013xfr} around the $D\bar D^*$ threshold, the $Z_c(4020)^\pm$~\cite{Ablikim:2013wzq,Ablikim:2013emm} near the $D^*\bar D^*$ threshold, the $Z_b(10610)^\pm$ and $Z_b(10650)^\pm$~\cite{Belle:2011aa,Garmash:2015rfd} near the $B\bar B^*$ and $B^*\bar B^*$ thresholds, the $Z_{cs}(3985)^-$~\cite{Ablikim:2020hsk} near the $\bar D_s D^*$ and $\bar D_s^* D$ thresholds, the $P_c$ states~\cite{Aaij:2019vzc} near the $\bar D^{(*)}\Sigma_c$ thresholds, and the dip and peak structures in the double $J/\psi$ spectrum~\cite{Aaij:2020fnh} near the $J/\psi\psi(2S)$ and $J/\psi\psi(3770)$ thresholds, respectively. 
These structures were suggested to be associated with threshold cusps~\cite{Bugg:2004rk,Bugg:2011jr,Chen:2011pv,Chen:2013coa,Swanson:2014tra,Swanson:2015bsa,Ikeda:2016zwx,Kuang:2020bnk,Sombillo:2020ccg,Dong:2020nwy,Wang:2020wrp,Karliner:2020dta,Cao:2020gul} (for a review, see Ref.~\cite{Guo:2019twa}). Notice that a pronounced threshold cusp requires the existence of a nearby pole~\cite{Bugg:2008wu,Guo:2014iya}.
However, so far it is not known when and at what thresholds, a nontrivial structure will show up.
In this Letter, we aim to answer this question. We will discuss the condition to have a peak at threshold, and show that it is natural to expect structures in the final state of a heavy quarkonium and light hadrons at the threshold of a pair of open-heavy-flavor hadrons that have attractive interaction at threshold.

{\it Single channel.}---Let us start with the effective range expansion for the $S$-wave amplitude of a two-body scattering (see Refs.~\cite{Guo:2017jvc,Brambilla:2019esw} for related discussions)
\begin{equation}
    f_0^{-1}(k) = \frac1{a_0} + \frac12r_0 k^2 - ik + \mathcal{O}\left(\frac{k^4}{\beta^4}\right),
\end{equation}
where $a_0$ and $r_0$ are the $S$-wave scattering length and effective range, $k$ is the magnitude of the c.m. momentum, and $\beta$ is some hard scale of the order of the inverse of force range. Here the sign convention for the scattering length is such that it is negative for repulsive and positive for attractive interaction in the absence of a bound state.
In the near-threshold region, we have the nonrelativistic expression for the momentum $k=\sqrt{2\mu E}$ with $\mu$ the reduced mass and $E$ the energy relative to the two-body threshold. One sees that the amplitude possesses a square-root branch point at $E=0$. As a result, there is a cusp exactly at the threshold in the modulus of the amplitude as a function of energy. 

If we focus on the region in the immediate vicinity of the threshold, we can neglect the effective range term. Writing the amplitude as a function of $E$, 
\begin{equation}
    f_0^{-1}(E) = \frac1{a_0} - i \sqrt{2\mu E},
    \label{eq:lo}
\end{equation}
we get
\begin{equation}
    |f_0(E)|^2 = \left\{
    \begin{array}{ll}
        \dfrac1{1/a_0^2 + 2\mu E} & \text{for } E\geq 0\\
        \dfrac1{\left(1/a_0 + \sqrt{-2\mu E}\right)^2} \quad & \text{for } E< 0
    \end{array}    
    \right. .
    \label{eq:abs0}
\end{equation}
It is easy to see that for positive $a_0>0$ (attractive interaction but not strong enough to form a bound state) this distribution is maximal, and thus has a cuspy peak, at the threshold $E=0$. Notice that in this case, there is a virtual state pole in the second Riemann sheet of the complex energy plane at $E_\text{virtual} = -1/(2\mu a_0^2)$.

In order to check what determines the shape of the cusp, we may compute the half-maximum width of $|f_0(E)|$. It is ready to be obtained from the solutions of $|f_0(E)|=|f_0(0)|/2$, $E_+ = \frac{3}{2\mu a_0^2}$ and $E_- = -\frac1{2\mu a_0^2}$,
and the half-maximum width is 
\begin{equation}
    \Gamma_\text{cusp} = E_+ - E_- = \frac{2}{\mu a_0^2} .
\end{equation}
One sees that the cusp in the energy distribution is narrower for a larger scattering length (thus stronger attraction) and also a larger reduced mass.
The former feature has been used to precisely measure the $S$-wave $\pi\pi$ scattering lengths~\cite{Pislak:2001bf,Pislak:2003sv,Batley:2007zz}.
The latter would imply that for a fixed scattering length, the signal of the threshold cusp becomes more and more evident when we increase the reduced mass, and it is natural to observe them in the heavy-hidden-flavor sector.

For a negative $a_0$, the distribution above threshold decreases monotonically in exactly the same way as the case of a positive $a_0$; below threshold, the maximum is located at the pole in the first Riemann sheet, $E_\text{bound} = -1/(2\mu a_0^2)$. 
For strong attraction, the pole is close to the threshold, and leads to a near-threshold peak.
For repulsive interaction, the value of $|a_0|$ is small compared to the range of forces; the pole is far away beyond the applicability region of the scattering length approximation (and thus is unphysical), and no nontrivial near-threshold structure exists.

Although we showed that the energy distribution of $|f_0(E)|$ has a maximum at the threshold for a positive $a_0$, the part below threshold needs to be observed in the final state of a lower channel. Thus, we need to consider a coupled-channel problem (see Ref.~\cite{Baru:2004xg} for an analysis of the Flatt\'e parametrization) to fully exhibit the threshold cusp structure.

{\it Coupled channels.}---Let us consider the energy region around the highest threshold in a coupled-channel system. Suppose that all open channels with lower thresholds are relatively far away so that the momentum variation for these channels around the highest threshold is smooth. In the spirit of the optical potential we may parametrize the scattering amplitude for the channel of interest in terms of a complex scattering length, i.e., Eq.~\eqref{eq:lo} with $a_0$ taking a complex value.
Unitarity requires $a_0$ to satisfy (see below)
\begin{equation}
    \im \frac1{a_0} \leq0 . \label{eq:unitarity}
\end{equation}
With a complex $a_0$, $|f_0(E)|^2$ becomes
\begin{align}
    \left\{
    \begin{array}{ll}
        \left[ \left(\re\frac1{a_0}\right)^2 + \left(\im \frac1{a_0} - \sqrt{2\mu E}\right)^2  \right]^{-1} & \text{for } E\geq 0\\
        \left[ \left(\im\frac1{a_0}\right)^2 + \left(\re \frac1{a_0} + \sqrt{-2\mu E}\right)^2  \right]^{-1} & \text{for } E< 0
    \end{array}    
    \right. .
    \label{eq:abs1}
\end{align}
One sees that the line shape decreases monotonically above threshold because of the unitarity constraint, Eq.~\eqref{eq:unitarity}, and it also decreases below threshold if the real part of $1/a_0$ is positive, or $\re(a_0)>0$. The half-maximum width of $|f_0(E)|$ in this case is
\begin{equation}
    \frac1{\mu} \left(\frac4{|a_0|^2} - \sum_x x\sqrt{\frac3{|a_0|^2} + x^2}  \right),
\end{equation}
where the sum runs over $x = \im (1/a_0)$ and $\re(1/a_0)$.
Again the cusp width is inversely proportional to the reduced mass.

To be more specific, let us consider a two-channel (denoted by channel 1 and channel 2) problem  and focus on the energy region around the threshold of channel 2, which is the higher one, such that $|E|<\Delta$ with $E$ the c.m. energy relative to the higher threshold and $\Delta$ the difference between the two thresholds. One is ready to construct a nonrelativistic effective field theory (NREFT).
The c.m. momentum of channel 2 is given by
\begin{equation}
    k_2 = \sqrt{2\mu_2 E}.
\end{equation}
The c.m.momentum of channel 1 can be expanded in power series of $E$, corresponding to even powers of $k_2$, as
\begin{align}
    k_1 =&\, \frac1{2\sqrt{s}}\sqrt{[s-(m_{1,1}+m_{1,2})^2][s-(m_{1,1}-m_{1,2})^2]  } \nonumber\\
    =&\, \frac1{2\Sigma_2} \sqrt{\Delta(\Delta+2m_{1,1})(\Delta + 2m_{1,2})(\Sigma_1+\Sigma_2)  } 
    + \mathcal{O}(E), \label{eq:k1}
\end{align}
where $m_{i,1}$ and $m_{i,2}$ are the masses of particles in channel-$i$, $\Sigma_1$ and $\Sigma_2$ are the thresholds of the lower and higher channels, respectively, and $\sqrt{s}=\Sigma_2+E$.
At leading order (LO) of the $E$ expansion, the Bethe-Salpeter integral equation becomes an algebraic equation, and the $T$ matrix can be written as
\begin{equation}
    T(E) = \left[1/V^\Lambda - G^\Lambda(E) \right]^{-1},
    \label{eq:lse}
\end{equation}
where $G^\Lambda(E)$ is a diagonal matrix containing the Green's functions for both channels. Around the channel-2 threshold, we treat the propagators of particles in this channel nonrelativistically, while those for the lower channel may be kept relativistically. That is,
\begin{align}
    G^\Lambda_1(E) = &\, i \int^{\Lambda_1}\!\!\! \frac{d^4q}{(2\pi)^2} \frac1{(q^2-m_{1,1}^2+i\epsilon)[(P-q)^2-m_{1,2}^2+i\epsilon]} \nonumber\\
    = &\, R(\Lambda_1) - i \frac{k_1}{8\pi\sqrt{s}},\\  
    G^\Lambda_2(E) = &\, -\frac{1}{4m_{2,1}m_{2,2}} \int^{\Lambda_2} \frac{d^3 \mathbf{q}}{(2\pi)^3} \frac{2\mu_2}{\mathbf{q}^2 - 2\mu_2 E-i\epsilon} \nonumber\\
    = &\, \frac{1}{8\pi \Sigma_2} \left[-\frac{2\Lambda_2}{\pi} + \, \sqrt{-2\mu_2 E-i\epsilon} + \mathcal{O}\left(\frac{k_2^2}{\Lambda_2}\right) \right],
\end{align}
with $P^2=s$ and $\mu_2$ the reduced mass of channel 2. Both loop integrals are ultraviolet divergent, and need to be regularized. Here, the nonrelativistic loop has been regularized using a hard cutoff $\Lambda_2>k_2$, for which higher order terms of $\mathcal{O}(k^2/\Lambda_2^2)$ will be neglected, while a different regularization method may be applied to the relativistic one. For the latter, we do not show the explicit expression, but notice that the real part, $R(\Lambda_1)$, can be expanded in Taylor series of $E$. 
Although $k_1$ can also be expanded as in Eq.~\eqref{eq:k1}, we keep it explicitly as it is regulator independent.

At LO, $V^\Lambda$ is a symmetric constant matrix, so does its inverse. The regulator-dependent terms in $G_1^\Lambda$ and $G_2^\Lambda$ can be renormalized by the diagonal matrix elements of $1/V^\Lambda$. Hence, we can write the $T$ matrix, which is cutoff independent, as
\begin{align}
    T(E) =&\, 8\pi \Sigma_2 \begin{pmatrix}
        -\frac1{a_{11}} + i k_1 & \frac1{a_{12}} \\[1mm]
        \frac1{a_{12}} & -\frac1{a_{22}} - \sqrt{-2\mu_2 E -i\epsilon}
    \end{pmatrix}^{-1} \nonumber\\
    =& -\frac{8\pi \Sigma_2}{\text{det}} 
    \begin{pmatrix}
        \frac1{a_{22}} + \sqrt{-2\mu_2 E-i\epsilon} & \frac1{a_{12}} \\
        \frac1{a_{12}} & \frac1{a_{11}} - i k_1
    \end{pmatrix},
    \label{eq:tmatrix}
\end{align}
with $\text{det} = \left(\frac1{a_{11}} - i\, k_1\right) \left(\frac1{a_{22}} + \sqrt{-2\mu_2 E-i\epsilon}\right) -\frac1{a_{12}^2}$. It is rather similar to the coupled-channel NREFT constructed in Ref.~\cite{Cohen:2004kf} (see also Ref.~\cite{Braaten:2005jj}), where both channels are treated nonrelativistically.
When the channel-coupling parameter $1/a_{12}$ vanishes, the system is reduced to two single channels.

Denoting the scattering length for the higher channel including the effects from the open channel as $a_{22,\text{eff}}$, one has
\begin{equation}
    \frac1{a_{22,\text{eff}}} = \frac1{a_{22}} - \frac{a_{11}}{a_{12}^2(1 + a_{11}^2 k_1^2)}  - i \frac{a_{11}^2k_1}{a_{12}^2(1+a_{11}^{2}k_1^{2} )}.
\end{equation}
The statement in Eq.~\eqref{eq:unitarity} is proven.

\begin{figure}[tb]
    \centering
    \includegraphics[width=\linewidth]{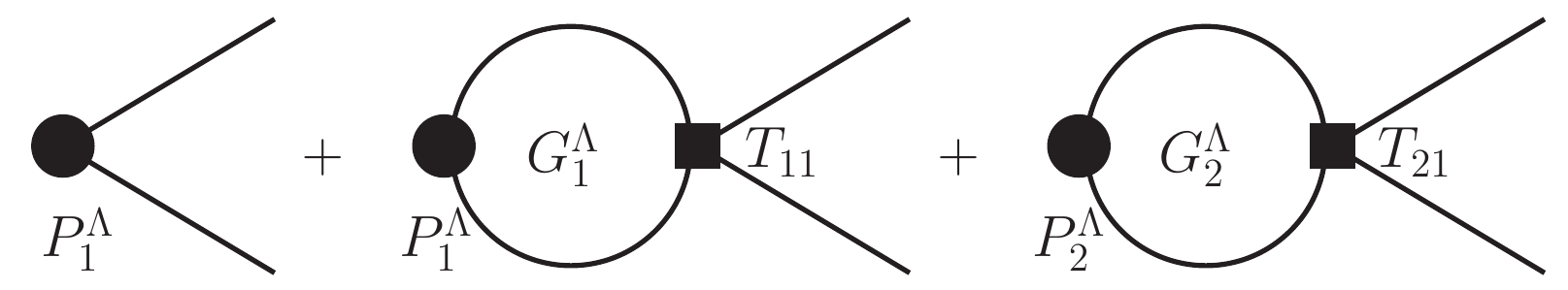}
    \caption{Production of particles in channel 1 through intermediate states in both channels.}
    \label{fig:prod}
\end{figure}

For the production of the two particles in channel 1 in the energy region close to the threshold of channel 2 (see Fig.~\ref{fig:prod}), the amplitude may be written as
\begin{align}
    &\, P_1^\Lambda [1 + G_1^\Lambda T_{11}(E)] + P_2^\Lambda G_2^\Lambda(E) T_{21}(E) \nonumber\\
    =&\, P_1^\Lambda (V_{11}^\Lambda)^{-1} T_{11}(E) + \left[P_1^\Lambda (V_{11}^\Lambda)^{-1}V_{12}^\Lambda + P_2^\Lambda \right] G_2^\Lambda T_{21}(E)\nonumber\\
    \equiv&\, P_1 T_{11}(E) + P_2 T_{21}(E),
    \label{eq:prod}
\end{align}
where $V_{11,12}^\Lambda $ are the elements of the $V^\Lambda$ matrix in Eq.~\eqref{eq:lse}, and the short-distance factors $P_{1,2}$ are $\Lambda$-independent, analytic in $E$ and can be taken as constants at LO in the $E$ expansion. 
For the second term with $G_2^\Lambda$, we have used $\Lambda_2 > k_2$, and kept only the cutoff term in $G_2^\Lambda$; it gets multiplicatively renormalized by the prefactor $P_1^\Lambda (V_{11}^\Lambda)^{-1}V_{12}^\Lambda + P_2^\Lambda$ which should scale as $1/\Lambda_2$~\cite{Braaten:2005jj}. The final expression is cutoff independent.

From analyzing the $E$ dependence of the modulus squared of the amplitude in Eq.~\eqref{eq:prod}, we can deduce the behavior of the invariant mass distribution of particles in the lower channel around the higher threshold. We discuss two cases in the following. 
For illustration, we will show some line shapes for a system with the lower channel being the $J/\psi \pi^-$ and the higher one being the $D^0 D^{*-}$. [Note that we only use their masses. 
For analyses of the system related to the $Z_c(3900)$ considering more complicated dynamics, see Refs.~\cite{Wang:2013cya,Albaladejo:2015lob,Pilloni:2016obd,Gong:2016jzb,Danilkin:2020kce}.]

\begin{figure}[tb]
    \centering
    \includegraphics[width=\linewidth]{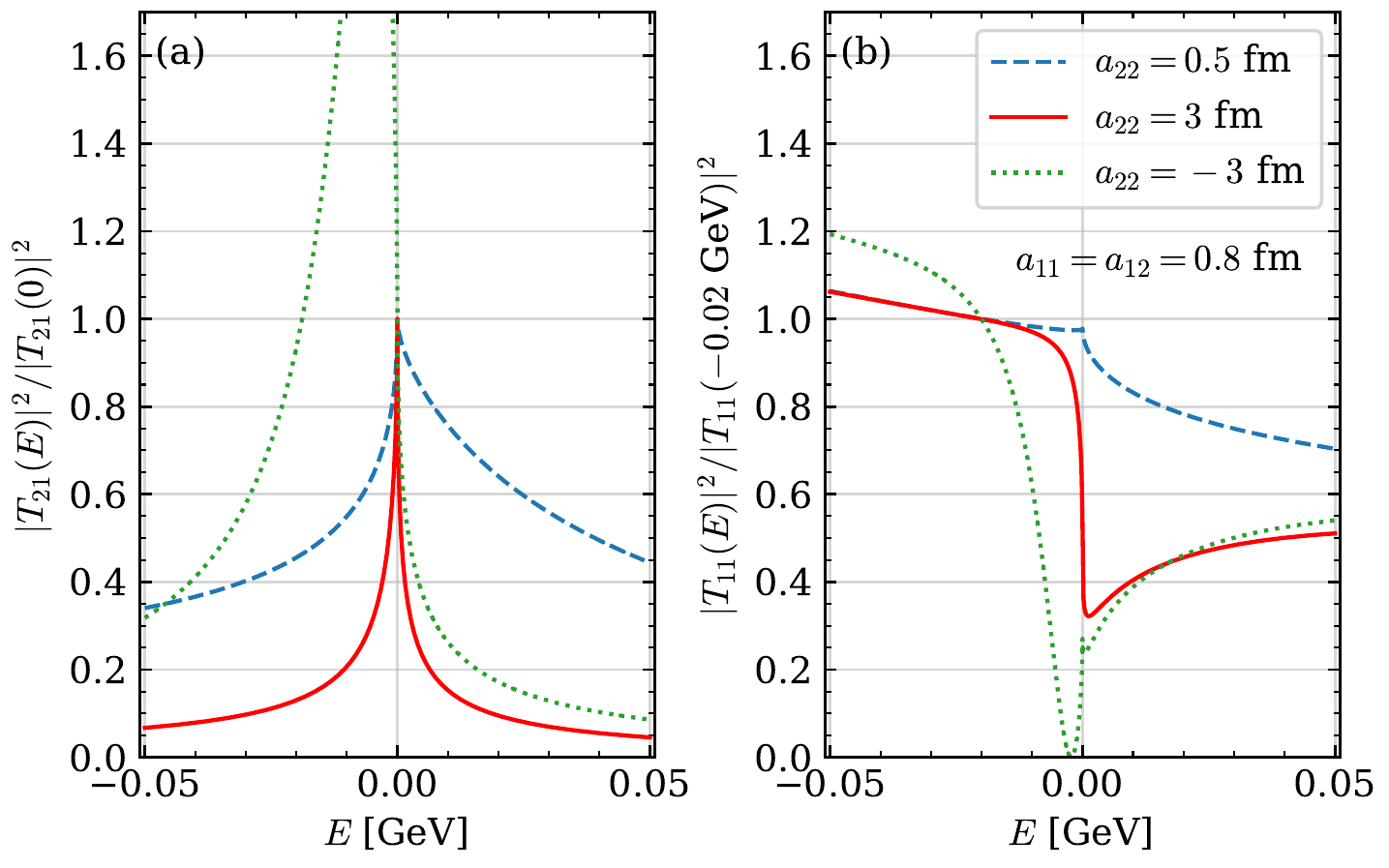}
    \caption{Illustration of threshold behaviors. Here we use the masses of the $\pi^-$ and $J/\psi$ for channel 1 and those of the $D^0$ and $D^{*-}$ for channel 2, and the values of used $a_{ij}$ parameters are given in the legend. (a) line shapes of $|T_{21}|^2$, which are normalized at the threshold of channel 2, i.e., $E=0$; (b) line shapes of $|T_{11}|^2$, which are normalized at $E=-0.02$~GeV.}
    \label{fig:cusps}
\end{figure}

Case 1: The production rate of channel 2 is much larger than that of channel 1, so that the production of channel 1 mainly proceeds through the last diagram in Fig.~\ref{fig:prod}, i.e., channel 2 is the driving channel. The event distribution is proportional to $|T_{21}(E)|$. We have
\begin{align}
    T_{21}(E) = \frac{-8\pi \Sigma_2}{a_{12}(1/a_{11}-i k_1)} \left[\frac1{a_{22,\text{eff}}} - i \sqrt{2\mu_2 E} + \mathcal{O}(E) \right]^{-1} .
    \label{eq:t21}
\end{align}
Since the only nontrivial energy dependence comes from the nonanalytic, second piece in the square brackets in Eq.~\eqref{eq:t21}, the analysis of Eq.~\eqref{eq:abs1} with a complex scattering length applies. Thus, the distribution maximizes at the higher threshold in its vicinity if the real part of $a_{22,\text{eff}}$ is positive.
This happens if the interaction for channel 2---which contains two parts: direct interaction in channel 2 and corrections from the channel coupling---is attractive, but not sufficient to produce a bound state. 
The red solid and blue dashed curves in Fig.~\ref{fig:cusps} (a) illustrate such a situation. The corresponding poles in the complex $k_2$ plane are at $(0.04-i\,0.08)$~GeV and $(0.37-i\,0.08)$~GeV, respectively.
When the pole is closer to the threshold of channel 2, the cusp is sharper.

It could happen that $a_{22}>0$ is large and $a_{11}$ is positive so that the channel coupling may render $\re(a_{22,\text{eff}})$ negative, or $a_{22}<0$ and $\re(a_{22,\text{eff}})$ remains negative after the channel coupling. In that case, the threshold would not be a local maximum, and there would be a peak below the higher threshold due to a below-threshold pole. 
An illustration is shown as the green dotted curve in Fig.~\ref{fig:cusps}~(a). The pole corresponding to these parameters is at $(-0.09-i\,0.08)$~GeV in the complex $k_2$ plane.

The effective scattering length $a_{22,\text{eff}}$ can be extracted using Eq.~\eqref{eq:t21} from a precise measurement of the threshold structure in this case.

Case 2: The production rate of channel 1 is much larger than that of channel 2, so that the production of channel 1 mainly proceeds through the first two diagrams in Fig.~\ref{fig:prod}. Then, the energy dependence of the production amplitude is dominated by that of $T_{11}(E)$. We have 
\begin{eqnarray}
    T_{11}(E) = \frac{-8\pi \Sigma_2 \left(\frac1{a_{22}} -i \sqrt{2\mu_2 E} \right) }{\left(\frac1{a_{11}} - i\, k_1\right) \left[\frac1{a_{22,\text{eff}}} -i \sqrt{2\mu_2 E} + \mathcal{O}(E) \right] }. ~~~~~
    \label{eq:t11}
\end{eqnarray}
One sees that if the channel coupling is weak (i.e., $a_{12}^2$ is large), the energy dependence around the higher threshold will be weak. 
However, if the channel coupling is strong, there will be nontrivial energy dependence.
For $E<0$, 
\begin{align}
    \frac{|T_{11}|}{8\pi\Sigma_2} = \left[k_1^2 + \left( \frac1{a_{11}} -  \frac1{a_{12}^2(1/a_{22} +\sqrt{-2\mu_2 E})} \right)^2 \right]^{-1/2}.
\end{align}
Thus, for $a_{22}>0$ and $1/a_{11} \leq a_{22}/a_{12}^2$, $|T_{11}(E)|$ increases monotonically when $E$ decreases from 0 (the red solid curve in Fig.~\ref{fig:cusps}~(b) presents an example).
For $E\geq 0$, 
\begin{align}
    \frac{|T_{11}|}{8\pi\Sigma_2} =&\, \left[ \left(\frac1{a_{11}} - \frac{1/(a_{22}a_{12}^2)}{1/a_{22}^2+2\mu_2 E}\right)^2 \right. \nonumber\\
    & \left. + \left(k_1 + \frac{\sqrt{2\mu_2 E}/a_{12}^2}{1/a_{22}^2 + 2\mu_2 E}\right)^2 \right]^{-1/2}.
\end{align}
The second term in the square brackets maximizes at $E = 1/(2\mu_2 a_{22}^2)$, while the first decreases for increasing $E$ when $a_{22}>0$ and $1/a_{11}\leq a_{22}/a_{12}^2$. Thus, whether $|T_{11}(E)|$ increases or decreases above the threshold of channel 2 depends on the competition between the two terms, and there must be a dip for a large $|a_{22}|$ because $T_{11}$ has a zero at $\sqrt{2\mu_2 E}=-i/a_{22}$, see Eq.~\eqref{eq:t11}.

In Fig.~\ref{fig:cusps}~(b), we show a few typical curves for the threshold behavior of $|T_{11}|^2$.
They are clearly different from those of $|T_{21}|^2$ shown in the left panel of the same figure. Although there is not necessarily a visible cusp or peak, there must be a dramatic change in the near-threshold region if there is a nearby pole in the amplitude, as can be seen from the red solid and green dotted curves. 
It is interesting to notice that the near-threshold behavior can be a sudden but continuous decrease around the threshold in the line shape.
Such a behavior is present around the $p\bar p$ threshold in the $\eta'\pi^+\pi^-$ invariant mass distribution measured by BESIII~\cite{Ablikim:2016itz}.

There is an important implication to the production mechanism for a near-threshold state which is rooted in channel 2 (large $a_{22}$): if it is observed as a peak in the final state of a lower channel, the driving production channel should be channel 2.
Let us consider the $f_0(980)$ and the $K\bar K$ threshold as an example. In the $J/\psi\to \phi\pi^+\pi^-$ and $J/\psi\to \omega\pi^+\pi^-$ processes, the $f_0(980)$ should be mainly from the $s\bar s$ and $(u\bar u+d\bar d)$ sources, respectively, and thus the $K\bar K$ (channel-2) and $\pi\pi$ (channel-1) as the corresponding meson channels.
The driving component for the production should be $T_{21}$ and $T_{11}$, respectively. 
Consequently, around the $K\bar K$ threshold, there is a narrow peak in the $\pi^+\pi^-$ distribution for the $J/\psi\to \phi\pi^+\pi^-$~\cite{Ablikim:2004wn} and a dip around the $K\bar K$ threshold for the $J/\psi\to \omega\pi^+\pi^-$~\cite{Ablikim:2004qna}. 
The analysis further supports that the driving channels for producing the $P_c$ peaks in $\Lambda_b$ decays are $\Sigma_c^{(*)}\bar D^{(*)}$~\cite{Liu:2015fea,Du:2019pij}, instead of $J/\psi p$~\cite{Roca:2015dva} or $\Lambda_c\bar D^{(*)}$~\cite{Burns:2019iih}.

More complex line shapes can be produced by the interference between the two terms in Eq.~\eqref{eq:prod}, i.e., between case 1 and case 2, if they have comparable strengths.

Many of the near-threshold structures have been measured in final states of the threshold channels, such as the $Z_c(3900)$ in $D\bar D^*$~\cite{Ablikim:2013xfr}, $Z_c(4020)$ in $D^*\bar D^*$~\cite{Ablikim:2013emm}, $Z_{cs}(3985)$ in $\bar D_s D^*+\bar D_s^* D$~\cite{Ablikim:2020hsk}, $Z_b(10610)$ in $B\bar B^*$~\cite{Garmash:2015rfd}.
The production amplitude should be 
\begin{equation}
    P_{2,\text{tot}}(E)=P_1 T_{12}(E) + P_2 T_{22}(E),
\end{equation}
with $P_{1,2}$ the short-distance parts. 
The event distribution in channel 2 is proportional to $k_2|P_{2,\text{tot}}(E)|^2$.
Since $1/a_{11}-i\,k_1$ is smooth around the threshold of channel 2, the $E$ dependence of $|T_{22}(E)|$ should be almost the same as that of $T_{12}(E)=T_{21}(E)$, see Eq.~\eqref{eq:tmatrix}, and thus the analysis of case 1 above applies.
In the final state of channel 2, only the $E\geq0$ piece can show up, which means that there should always be a near-threshold enhancement no matter the sign of $\re(a_{22,\text{eff}})$, see Eq.~\eqref{eq:abs1}.
However, the $k_2$ factor from the phase space would suppress the threshold behavior if the pole is not sufficiently close to the threshold.
This, as well as the analysis below Eq.~\eqref{eq:t21}, supports the conclusion in Ref.~\cite{Guo:2014iya} in an analysis of the $Z_c(3900)$ that a near-threshold prominent cusp implies a near-threshold pole.

{\it Discussions and conclusion.---}From the discussions above, one expects that there should always be a peak around the threshold for a two-body system with attractive interaction in the invariant mass distribution of a lower channel (channel-1), if the production proceeds mainly through the channel with the relevant threshold, i.e., channel 2 above.
This should apply to all hadron pairs with one containing a heavy quark and the other containing a heavy antiquark.
Because for the heavy quark-antiquark pair to form a heavy quarkonium, the relative momentum should be small, and it should only be a small part of the whole phase space, the production of a pair of open-heavy-flavor hadrons should be much easier than that for a heavy quarkonium plus light hadrons. 
Consequently, it is natural to expect that the mechanism shown as the last diagram in Fig.~\ref{fig:prod} should be significant for the production of a heavy quarkonium plus light hadrons when open-flavor thresholds are open, and there should be a near-threshold peak if the interaction in the open-flavor channel is attractive. The peak is a threshold cusp if  $\re(a_{22,\text{eff}})$ is positive, and is a peak just below threshold if the attraction is strong enough to produce a below-threshold bound state such that $\re(a_{22,\text{eff}})$ becomes negative and its absolute value is large.
It is interesting to notice that with the same effective scattering length, the larger the reduced mass is, the sharper the peak would be.

Notice that the threshold behavior gets smeared if at least one of the relevant intermediate particles carries a finite width. Hence, it should be prominent only for those with negligible or tiny widths.
The existence of a triangle singularity in special cases, when the kinematics is such that the singularity is close to the physical region, can lead to additional nontrivial structure at or above threshold (for a detailed discussion, see the review~\cite{Guo:2019twa}).

In Table~\ref{tab:attr}, we list heavy-antiheavy hadron pairs (taking the charmed ones for example) that are expected to have attractive interaction at threshold.
The conclusion is based on the one-boson exchange model with Lagrangians constructed considering heavy quark spin symmetry~\cite{Yan:1992gz,Casalbuoni:1996pg,Liu:2011xc}.
The pseudoscalar exchanges in the chiral limit always yield potentials proportional to the square of transferred momentum, ${\bm q}^2$, which vanish at threshold. Hence, only vector-meson exchanges are considered, which is analogous to the vector-meson dominance contribution to the low-energy constants in chiral perturbation theory~\cite{Ecker:1988te} (see the Supplemental Material for a list of the potentials).
Calculations of heavy-antiheavy hadron interactions using lattice QCD are indispensable to reach model-independent results on which hadron pairs are attractive, and thus to understanding whether there should be a nontrivial structure around a given threshold.

It is expected that near-threshold peaks for these hadron pairs will show up in final states of a heavy quarkonium plus light hadrons. One also notices that the Born term contribution from the light-boson exchange to the scattering length scales as the heavy quark mass $m_Q$, and one would expect it stronger in the bottom-antibottom than in the charm-anticharm sector.

The predictions made here are ready to be tested, and the threshold structures will play an important role in revealing the interactions between a pair of heavy hadrons, and thus to the mysterious hadron structures with a pair of heavy quark and antiquark.
At last, the NREFT analysis of the threshold behavior is general enough and may find its applications in fields other than hadron physics.

\begin{table}
\caption{Charm-anticharm hadron pairs that have attractive interaction at threshold from vector-meson exchanges (similar in the bottom sector). We use $H, T$ and $S$ to denote the ground state heavy mesons, SU(3) antitriplet and sextet heavy baryons, respectively. Isospin is labelled for each pair in square brackets.
Those with $^\dag$ mean that the contribution from the light-vector exchanges vanishes, and the attraction is provided by sub-leading exchanges of vector charmonia~\cite{Aceti:2014uea}. } \label{tab:attr}
\renewcommand*{\arraystretch}{1.25}
\begin{ruledtabular}
\begin{tabular}{l| l l l}
$H\bar H$& $D^{(*)}\bar D^{(*)} [0,1^\dag]$;& $D_s^{(*)}\bar D^{(*)}$ $[\frac12^\dag]$;&
$D^{(*)}_s\bar D^{(*)}_s\,[0]$\\\hline
$\bar H T$ & $\bar D^{(*)}\Xi_c\,[0]$;& $\bar D_s^{(*)}\Lambda_c\,[0^\dag]$\\\hline
$\bar H S$ & $\bar D^{(*)}\Sigma_c^{(*)}\,[\frac12]$;& $\bar D_s^{(*)}\Sigma_c^{(*)}\,[1^\dag]$;& $\bar D^{(*)}\Xi_c^{\prime(*)}\,[0]$; \\
& $\bar D^{(*)}\Omega_c^{(*)}\,[\frac12^\dag]$ \\\hline
$T\bar T$ & $\Lambda_c\bar \Lambda_c\,[0]$; & $\Lambda_c\bar \Xi_c\, [\frac12]$;& $\Xi_c\bar \Xi_c\, [0,1]$ \\\hline
$T\bar S$ & $\Lambda_c\bar\Sigma_c^{(*)}\,[1]$;& $\Lambda_c\bar\Xi_c^{'(*)}\,[\frac12]$;& $\Lambda_c\bar\Omega_c^{(*)}\, [0^\dag]$;\\
& $\Xi_c \bar\Sigma_c^{(*)}\, [\frac32^\dag, \frac12]$;
& $\Xi_c \bar\Xi_c^{'(*)}\,[1,0]$;& $\Xi_c \bar\Omega_c^{(*)}\,[\frac12]$\\\hline
$S\bar S$ & $\Sigma_c^{(*)}\bar\Sigma_c^{(*)}\,[2^\dag,1,0]$; & 
$\Sigma_c^{(*)}\bar\Xi^{'(*)}_c\,[\frac32^\dag,\frac12]$;& $\Sigma_c^{(*)}\bar\Omega^{(*)}_c\, [0^\dag]$;\\
& $\Xi_c^{'(*)} \bar\Xi_c^{'(*)}\,[1,0]$;& $\Xi^{'(*)}_c \bar\Omega_c^{(*)}\,[\frac12]$;& $\Omega_c ^{(*)}\bar\Omega_c^{(*)}\,[0]$
\end{tabular}
\end{ruledtabular}
\end{table}

\vspace{0.5cm}

\begin{acknowledgments}
FKG is grateful to Zhi-Hui Guo, Christoph Hanhart and Su Yi for helpful discussions.
This work is supported in part by the National Natural Science Foundation of China (NSFC) under Grants No.~11835015, No.~12047503 and No.~11961141012, by the NSFC and the Deutsche Forschungsgemeinschaft (DFG, German Research Foundation) through the funds provided to the Sino-German Collaborative Research Center TRR110 ``Symmetries and the Emergence of Structure in QCD" (NSFC Grant No.~12070131001, DFG Project-ID 196253076), by the Chinese Academy of Sciences (CAS) under Grant No.~XDB34030000 and No.~QYZDB-SSW-SYS013, and by the CAS Center for Excellence in Particle Physics (CCEPP).
\end{acknowledgments}

\bibliographystyle{apsrev}
\bibliography{cusps}

\section*{Supplemental Material}

Considering the exchange of vector mesons, the potential between a pair of heavy and antiheavy hadrons at threshold takes the following form: 
\begin{equation}
    V\sim-F\beta_1\beta_2g_V^2\frac{2m_1m_2}{m_{\rm ex}^2},
    \label{eq:potential}
\end{equation}
where $m_1,m_2$ and $m_{\rm ex}$ are the masses of the two heavy hadrons and the exchanged particle, respectively, $\beta_1$ and $\beta_2$ are the coupling constants for the two heavy hadrons with vector mesons, $g_V$ is a coupling parameter for the light-vector mesons, and $F$ is a group theory factor accounting for light-flavor SU(3) information. The values of $F$ are listed in Table~\ref{tab:potentials} for all combinations of a pair of heavy and antiheavy ground state hadrons. $\beta_1$ and $\beta_2$ are positive in our convention so that a positive $F$ means an attractive interaction. For systems that can form states with both positive and negative $C$ parities, for instance, $D\bar D^*\pm\bar D D^*$ or $\Sigma_c \bar \Sigma_c^*\pm \bar \Sigma_c \Sigma_c^*$, the potentials at threshold are the same with the mechanism considered here. The potentials presented here may also be used as the resonance saturation modeling of the constant contact terms in nonrelativistic effective field theory studies of the heavy-antiheavy hadron interactions.

\begin{table}[h]
\caption{Potentials at threshold of heavy-antiheavy hadron pairs with only light vector-meson exchanges, see Eq.~\eqref{eq:potential}. Positive $F$ means attractive. For the systems with $F=0$, the sub-leading exchanges of  vector-charmonia also lead to an attractive potential at threshold. } \label{tab:potentials}

\begin{ruledtabular}
\begin{tabular}{cccc}
System &  $I$ & exchanged particle & $F$\\
\hline
$D^{(*)}\bar D^{(*)}$& 0 &$\rho,\omega$ & $\frac32,\frac12$\\
& 1 &$\rho,\omega$ & $-\frac12,\frac12$\\
$D_s^{(*)}\bar D^{(*)}$& $\frac12$ &$-$ & $0$\\
$D^{(*)}_s\bar D^{(*)}_s $& 0&$\phi$ & $1$\\
\hline
$\bar D^{(*)}\Lambda_c$& $\frac12$ &$\omega$ & $-1$\\
$\bar D_s^{(*)}\Lambda_c$& $0$ &$-$ & $0$\\
$\bar D^{(*)}\Xi_c$& $1$ &$\rho,\omega$ & $-\frac12,-\frac12$\\
& $0$ &$\rho,\omega$ & $\frac32,-\frac12$\\
$\bar D_s^{(*)}\Xi_c$& $\frac12$ &$\phi$ & $-1$\\
\hline
$\bar D^{(*)}\Sigma_c^{(*)}$& $\frac32$ &$\rho,\omega$ & $-1,-1$\\
& $\frac12$ &$\rho,\omega$ & $2,-1$\\
$\bar D_s^{(*)}\Sigma_c^{(*)}$& $1$ &$-$ & $0$\\
$\bar D^{(*)}\Xi_c^{'(*)}$& $1$ &$\rho,\omega$ & $-\frac12,-\frac12$\\
& $0$ &$\rho,\omega$ & $\frac32,-\frac12$\\
$\bar D_s^{(*)}\Xi_c^{'(*)}$& $\frac12$ &$\phi$ & $-1$\\
$\bar D^{(*)}\Omega_c^{(*)}$& $\frac12$ &$-$ & $0$\\
$\bar D_s^{(*)}\Omega_c^{(*)}$& $0$ &$\phi$ & $-2$\\
\hline
$ \Lambda_c\bar\Lambda_c$& $0$ &$\omega$ & $2$\\
$\Lambda_c\bar \Xi_c$& $\frac12$ &$\omega$ & $1$\\
$\Xi_c\bar \Xi_c$& $1$ &$\rho,\omega,\phi$ & $-\frac12,\frac12,1$\\
& $0$ &$\rho,\omega,\phi$ & $\frac32,\frac12,1$\\
\hline
$\Lambda_c\bar\Sigma_c^{(*)}$& $1$ &$\omega$ & $2$\\

$\Lambda_c\bar\Xi_c^{'(*)}$&$\frac12$ &$\omega$ & $1$\\

$\Lambda_c\bar\Omega_c^{(*)}$ &$0$ &$-$ & $0$\\

$\Xi_c \bar\Sigma_c^{(*)}$  &$\frac32$ &$\rho,\omega$ & $-1,1$\\

&$\frac12$ &$\rho,\omega$ & $2,1$\\
$\Xi_c \bar\Xi_c^{'(*)}$ &$1$ &$\rho,\omega,\phi$ & $-\frac12,\frac12,1$\\

& $0$ &$\rho,\omega,\phi$ & $\frac32,\frac12,1$\\
$\Xi_c \bar\Omega_c^{(*)}$ &$\frac12$ &$\phi$ & $2$\\
\hline
$\Sigma_c^{(*)}\bar\Sigma_c^{(*)}$  & $2$ &$\rho,\omega$ & $-2,2$\\

 & $1$ &$\rho,\omega$ & $2,2$\\

 & $0$ &$\rho,\omega$ & $4,2$\\

$\Sigma_c^{(*)}\bar\Xi^{'(*)}_c$  &$\frac32$ &$\rho,\omega$ & $-1,1$\\

 & $\frac12$ &$\rho,\omega$ & $2,1$\\

$\Sigma_c^{(*)}\bar\Omega^{(*)}_c$  &$0$ &$-$ & $0$\\

$\Xi_c^{'(*)} \bar\Xi_c^{'(*)}$&$1$ &$\rho,\omega,\phi$ & $-\frac12,\frac12,1$\\

 &$0$ &$\rho,\omega,\phi$ & $\frac32,\frac12,1$\\

$\Xi^{'(*)}_c \bar\Omega_c^{(*)}$&$\frac12$ &$\phi$ & $2$\\

$\Omega_c ^{(*)}\bar\Omega_c^{(*)}$ &$0$ &$\phi$ & $4$
\end{tabular}
\end{ruledtabular}
\end{table}

\end{document}